\documentstyle{article}
\title{A classical perspective on nonlocality in \\
         quantum field theory}
\author{P. W. Morgan\\
        {\small\sl \vbox{\vskip 3pt
                         \hbox{30, Shelley Road,} 
                         \hbox{Oxford, OX4 3EB,} 
                         \hbox{England.}}}\\
        {\small\sl \hbox{peter.morgan@philosophy.oxford.ac.uk}}}
\date{November 7th, 2001}
\def\Xabstract{A classical statistical field theory hidden variable model
for the quantized Klein-Gordon model is constructed that preserves
relativistic signal locality and is relativistically covariant, but is at
the same time relativistically nonlocal, paralleling the Hegerfeldt
nonlocality of quantum theory. It is argued that the relativistic
nonlocality of this model is acceptable to classical physics, but in any
case the approach taken here characterizes the nonlocality of the
quantized Klein-Gordon model in terms of concepts from classical statistical
field theory.}

\begin{document}
\hoffset=-0.130\hsize
\hsize = 1.2\hsize

\def\Planck{{\vrule width 0.4em height 0.53em depth -0.5em\kern -0.45em h}}
\def\Half{{\scriptstyle {\scriptstyle 1 \over \scriptstyle 2}}}
\def\kT{{{\rm\sl k}T}}
\def\RR{{\rm I\!R}}

\def\citeJammerP156{{Jammer(1966, p. 156)}}
\def\citeIZP119{{Itzykson and Zuber(1980, p. 119)}}
\def\citeBinneyS81{{Binney {\it et al.}(1993, \S 8.1)}}
\def\citeBinneyAL32{{Binney {\it et al.}(1993, Appendix L.3.2)}}
\def\citeHegerfeldt{{Hegerfeldt(1998)}}
\def\citeKaloyerou{{Kaloyerou(1996)}}
\def\citeHaag324{{Haag(1996, Theorem 3.2.4)}}
\def\citeHaagSII53{{Haag(1996, \S II.5.3)}}
\def\citeBuchholz{{Buchholz and Yngvason(1994)}}
\def\citeRedhead{{Redhead(1987, ch. 5)}}
\def\citeSG{{Segal and Goodman(1965)}}
\def\citeMorgan{{Morgan(2001)}}

\maketitle

\hsize=0.9\hsize
\hangindent=0.111\hsize\hangafter=-20
\Xabstract

\vskip 30pt
\hsize=1.1111\hsize
\baselineskip 14.5pt

\section{Introduction}
This paper takes a relativistically local classical model for quantum field
theory not to be possible. Obviously we then have the choice of abandoning
classical models or considering what relativistically nonlocal classical
models are possible. We take here a classical statistical field theory
approach, which leads to a model of classical probability measures over
classical fields for the quantized Klein-Gordon model that preserves
relativistic signal locality and is relativistically covariant despite
being relativistically nonlocal (that is, there are causal dynamical
relationships between classical field values at space-like separated
points).

We will pursue infinite dimensional classical statistical field theory
models for quantum field theory because I take it that ways of introducing
the necessary relativistic nonlocality into finite dimensional classical
models for non-relativistic quantum theory are not acceptable (and I can't
see any better way), and because models of quantum field theory are
empirically superior to finite dimensional models of non-relativistic
quantum mechanics. If relativistic nonlocality is acceptably introduced
into classical statistical field theory in a way that adequately matches
the Hegerfeldt nonlocality of quantum field theory, then an approximate
reduction of both a classical statistical field theory model and the
corresponding quantum field theory model to finite dimensional models will
result in a finite dimensional classical model that in general will not
perfectly describe the relativistic nonlocality present in either the
quantum field theory model or its finite dimensional reduction. The
relativistic nonlocality in such finite dimensional classical models may
not be acceptable as a fundamental approach, but it may nonetheless be
useful as the consequence of an approximate reduction of an acceptable
classical statistical field theory.

Although classical models can be taken realistically, I take them, in a
post-empiricist way, to be `as if' real. The construction of classical
models is often motivated by a na\"\i ve classical realism, but a
post-empiricist view is simply that it is useful to have available models
of many different types. Visualizability of a model is a pragmatic virtue,
and does not prove that the world is the same as such mathematics. Equally,
although the possibility of constructing models of great accuracy is an
important pragmatic virtue, it also does not prove that the world is
ultimately very like a model of a given theory. It may be only marginally
useful to have models that are visualizable in competition with the numerous
existing interpretations of quantum field theory, but I take it to be at
least a little useful.

In the absence of a convincing post-empiricist interpretation of classical
and quantum physics, section 2 is a brief outline of an approach to
classical and quantum models that is largely a polemic, but does have some
implications for the subsequent development, because it describes the
attitude that will be taken to the Kochen-Specker paradox. It can safely be
ignored by those who wish to rush to the mathematical construction of a
classical statistical field theory model for the quantized Klein-Gordon
model, which is contained in section 3. Finally, section 4 argues that the
relativistic nonlocality of this classical model is acceptable to classical
physics.

\section{On the interpretation of classical and quantum theory}
A classical statistical field theory model does not describe measurement
directly. If we introduce a device to measure the field somewhere, then
in principle we have to construct a bigger model which includes a model of
the device. This requires another device to measure properties of the
first device, {\it etc.}, {\it etc.} To `avoid' this regress (although it
remains as a formal problem), measurement in a classical statistical field
theory model requires a feature in the model to be pragmatically
identifiable in the world. The way in which we make this pragmatic
identification depends on the level of approximation we require. At a
quite direct level, a meter pointer in a model will be in some
configuration relative to other parts of the model, and we can see
objects in the world which we take to be in approximately the same
relative configuration. We assume that the meter pointer is not affected
much by us seeing it, so that we do not have to include ourselves in the
model. Less directly, if an electric current is large enough in a model
to be measured without significant perturbation by an ammeter, we do
not need to model a particular ammeter. Still less directly, and most
usually (and as we shall do in section 3), we may construct a classical
statistical field theory model quite abstractly, without any concessions
to measurement\footnote{Very often, classical statistical field theory
models occupy all of the space-time of the model; there is no room for
a measurement device (or for us) without making the model mathematically
quite intractable.}, and know that there are classes of functions on the
phase space that correspond with moderate inaccuracy and perturbation to
measurements of real finite systems by devices we have available to us. In
a classical statistical field theory model, the inaccuracy of and
perturbation caused by a measurement device is in principle supposed to be
arbitrarily reducible, but it is in principle as well as in practice that
we have only a finite array of imperfect devices, so not all details of a
classical model are verifiable. 

We will take quantum theory in a similar, quite pragmatic way. Quantum
theory is a formalism for summarizing the results of multiple experimental
arrangements, which we can largely take to consist of different arrangements
of carefully calibrated preparation devices and measurement devices. Our
description of a measurement device is quite pragmatic: we might describe a
device in a first approximation as measuring momentum, but refine our
description to a positive operator valued measure, which we can construct
partly from approximate classical measurements of the device and partly
through a finite process of calibration. We never know precisely what
observable we measure with a given real device. The regress of measurement
devices described above for classical models is of course very well known
in quantum theory, and can be `avoided' in a similarly pragmatic way.

Following the Copenhagen interpretation, each arrangement of preparation and
measurement devices is described classically, as are the results we obtain.
We take a quantum theory description of multiple experimental arrangements
to be a theoretical construction from many classical descriptions.
Consequently, we will require a classical statistical field theory model
only to give an account for a classical description of an individual
experiment, and not to give an account for the multiple classical
descriptions that are summarized by a quantum theory description.

We will find that Planck's constant can be considered to be a measure of
fluctuations of a classical field: we cannot improve on the Heisenberg
uncertainty principle because all our measurement devices are subject to
the same fluctuations. This is a very old understanding of quantum theory
in classical terms, which can only work adequately, however, if
relativistically nonlocal models are accepted. We take it that insofar as a
classical statistical field theory model has properties that can only be
measured using a device that improves upon a classical version of this
limit, those properties are in fact not verifiable, at least at present.
Unless we can find a way of reducing the fluctuations of the field locally,
we will of course never be able to improve on the Heisenberg uncertainty
principle.

There is a little used approach to understanding quantum mechanics, due to 
Mackey, that is more-or-less consistent with the classical statistical field
theory approach to quantum field theory taken here. For Mackey, quantum
mechanics is ``basically a revision of statistical mechanics in that one
studies the change in time of probability measures but no longer supposes
that the motion of these measures is that induced by a motion of points in
phase space''(quoted by \citeJammerP156). This is quite a conventional
statement for Mackey, which just asserts that there are no trajectories, but
we will apply it to quantum field theory as a starting point for a modal
interpretation. We will regard momentum information in both classical and
quantum mechanics as the additional information required to describe the
evolution of a probability measure over a configuration space of classical
fields, even though the additional information required is very different
in the two cases. Momentum information in quantum field theory is then a
concept particular to quantum field theory, which is not `real' in a
classical statistical field theory model, even though probability measures
over configuration space are taken to be a common conceptual foundation.
When we say in a quantum theory model that we measure quantum momentum (or
any other observable that is not compatible with the field observable) with
a given device, we can in principle construct a (relativistically nonlocal)
classical model of the whole apparatus in which the {\sl position} of a
classical pointer (say) is correlated with some function of the classical
state.

Although ideal classical measurement is noncontextual, each combination of a
classical statistical field theory model with a model of a real measurement
device changes the details of the classical statistical field theory model
in a different way, so the properties of the field, when considered in
combination with multiple measurement devices, are essentially contextual.
Both this contextuality and the approach of the last paragraph, taking 
quantum momentum as not a classical observable, are natural ways to evade
the strictures of the Kochen-Specker paradox (see, for example,
\citeRedhead). The contextuality of real measurement, in particular, is very
natural in a classical statistical field theory.

Mackey's approach leads to a modal interpretation of quantum field theory
that is different from a classical statistical field theory only because of
the way in which the evolution of the common probability measure over a
common sample space of classical field configurations is specified. Quantum
mechanical momentum describes the evolution of the probability measure over
classical field configurations directly, whereas classical momentum
describes the evolution of single classical field configurations, indirectly
inducing the evolution of the probability measure. A classical statistical
field theory, however, in general makes no attempt to compute individual
trajectories, but manipulates classical probability measures in a way that
is quite comparable to that of quantum field theory. Mackey's approach
applied to non-relativistic, finite-dimensional quantum mechanics is not
particularly helpful, largely because the violation of Bell inequalities
remains problematic in its implication of relativistic nonlocality, but the
classical relativistic nonlocality implicit in the evolution of probability
measures over a configuration space of classical fields described by a
quantum field theory can at least potentially be reasonable as a classical
thermal physics. Such an interpretation of quantum field theory is
non-relativistic, but of course the same interpretation is possible for any
observer with their own choice of configuration space; it is to a Lorentz
covariant interpretation as a Hamiltonian formalism is to a Lagrangian
formalism --- it is not manifestly covariant, but it is nonetheless
applicable.

As a modal interpretation of quantum field theory, the approach we have
taken is similar to Kaloyerou's Bohmian approach to the interpretation of
quantum field theory. At least for the observables of boson fields, and
in principle for the observables of fermion fields, \citeKaloyerou\ takes
quantum field theory to describe a probability measure over classical
fields. Mackey's approach differs from a Bohmian approach because it is
not committed to a particular classical evolution, and the Hamiltonian
ansatz introduced below is also quite different to introducing a Bohmian
nonlocal evolution. From the point of view both of classical statistical
field theory and of quantum field theory, it is in principle not possible
from experimental evidence to identify a Hamiltonian with a Bohmian degree
of precision, even if we believe or act as if there might be one.

\section{A classical relativistically nonlocal model for \vskip 2pt
         \hskip 10pt the quantized Klein-Gordon model}
In simple-minded terms, a classical statistical field theory at equilibrium
describes the consequences of a Gibbs probability measure over a sample
space of classical fields. For the simplest possible model, the Gaussian
model, with a Hamiltonian 
$$H[\phi] = \int\Half\{(\nabla\phi)^2 +m^2\phi^2\}d^3x,$$
and probability density $P[\phi]$ proportional to $\exp{\!(-H[\phi]/\kT)}$,
the fluctuations of the field are Gaussian distributed, with variance
$$\sigma^2 = \kT\int {d^3k\over(2\pi)^3}
                         {|\tilde f(k)|^2\over{k^2+m^2}},$$
where $f(x)$ is a test function used to construct a classical observable
$$\phi_f = \int\phi(x)f(x)d^3(x)$$
(see, for example, \citeBinneyS81). We have had to introduce the test
function $f(x)$ because the variance of the fluctuations `at a point' are
infinite, as we see if we take $f(x)$ to be a delta function, for which
$|\tilde f(k)|^2 = 1$. Provided we choose $\tilde f(k)$ appropriately,
$\sigma^2$ will be finite.

The Hamiltonian of the Gaussian model can be written in fourier transform
terms as
$$H[\phi] = \int{d^3k\over(2\pi)^3} \Half(k^2+m^2)|\tilde\phi(k)|^2.$$
If we consider a more general ansatz for a non-interacting Hamiltonian,
$$H_\xi[\phi] = \int{d^3k\over(2\pi)^3} \xi(k)|\tilde\phi(k)|^2,$$
generalizing the argument in \citeBinneyAL32, the fluctuations of the field
are again Gaussian distributed, with variance
$$\sigma^2_\xi = \kT\int {d^3k\over(2\pi)^3}
                         {|\tilde f(k)|^2 \over 2\xi(k)}.$$
Note that if $\xi(k)$ is chosen appropriately, the fluctuations at a
point are finite, and we can properly talk of a probability measure over
a sample space of classical fields. The simplest regularization of applying
a frequency cutoff corresponds to taking $\xi(k)=\Half(k^2+m^2)$ when
$k\leq\Lambda$, $\xi(k)=\infty$ when $k>\Lambda$, but smooth dynamical
regularizations can also be adopted, such as
$\xi(k) = \Half(k^2+m^2e^{|k|/\Lambda})$ or
$\xi(k) = \Half(k^2+m^2)^\alpha$, $\alpha>{3\over 2}$ (which is largely
equivalent to dimensional regularization).

We can compare the vacuum state of a quantized Klein-Gordon field, with 
Hamiltonian
$$\int\Half\{\hat\pi^2+(\nabla\hat\phi)^2 +m^2\hat\phi^2\}d^3x,$$
with the equilibrium state of the Hamiltonian ansatz $H_\xi$ in classical
statistical field theory. The formalisms of quantum field theory and of
classical statistical field theory are obviously very different, and
equilibrium in the Gaussian model is in general 3-dimensional Euclidean
invariant rather than 4-dimensional Poincar\'e invariant. Nonetheless,
fluctuations of the quantum field are Gaussian distributed in the quantized
Klein-Gordon vacuum state, just as they are in the Gaussian model, with
variance
$$\sigma^2_{QFT} = \Planck\int {d^3k\over(2\pi)^3}
                      {|\tilde f(k)|^2\over 2\sqrt{k^2+m^2}} =
                   \Planck\int {d^4k\over(2\pi)^4}2\pi
                      \delta(k^\mu k_\mu-m^2)\theta(k_0)|\tilde f(k)|^2$$
(see, for example, \citeIZP119), where again $f(x)$ is a test function used
to construct a quantum observable
$$\hat\phi_f = \int\hat\phi(x)f(x)d^4(x).$$
Although $f(x)$ can be any function on the whole of the 4-dimensional
Minkowski space, only components of the fourier transform $\tilde f(k)$
that are on-shell, satisfying $k^\mu k_\mu=m^2$, $k_0>0$, contribute to the
variance. Consequently, for this purpose any function is equivalent to any
other function that has the same on-shell components. In particular, $f(x)$
can be a product $f_{(3)}(x)\delta(t-t_0)$ of a function on a timelike
3-dimensional hyperplane with the delta function $\delta(t-t_0)$ which picks
out that hyperplane, allowing a direct comparison with a 3-dimensional
classical statistical field theory equilibrium state.

There is nothing absolutely quantum mechanical about the vacuum state of the
quantized Klein-Gordon field, because we can generate the same fluctuations
on a space-like hyperplane in a classical way by taking
$\xi(k) = \kT\sqrt{k^2+m^2}/\Planck$ (of course the $\kT$ just cancels with
the $\kT$ in the Gibbs distribution). These classical and quantum models
agree for all correlation functions that are restricted to a timelike
3-dimensional hyperspace: in both models, all connected correlation
functions are zero, except for
$$\left<\phi_f\phi_g\right> = \Planck\int {d^3k\over(2\pi)^3}
                         {\tilde f^*(k)\tilde g(k)\over 2\sqrt{k^2+m^2}}.$$
Within the Hamiltonian ansatz we have adopted, only
$\xi(k) = \kT\sqrt{k^2+m^2}/\Planck$ results in the same fluctuations as the
quantized Klein-Gordon vacuum state. Furthermore, a Gibbs probability measure
constructed using a simple perturbation of the Hamiltonian ansatz does not
result in the same fluctuations as the quantized Klein-Gordon state,
because in general such perturbations generate a non-Gaussian distributed
probability measure. So this relativistically nonlocal classical Hamiltonian
is the only candidate for a classical statistical field theory
hidden-variable model for the quantized Klein-Gordon model, unless a
significantly more complicated and less conventional ansatz is adopted. 

In a classical statistical field theory framework, we can understand
Planck's constant of action as a Poincar\'e invariant measure of field
fluctuations that is a 4-dimensional analogue of the thermal energy $\kT$,
which can itself be understood as a 3-dimensional Euclidean invariant
measure of field fluctuations. That is, Planck's constant can be regarded
as no longer a fundamental constant, but in a loose sense as a measure of
``4-dimensional temperature''.

In classical terms, a state in which there are fluctuations everywhere is
clearly not a minimum of energy. What is at a minimum for the vacuum state
compared to states in the Fock space, which are asymptotically identical to
the vacuum state, is the Helmholtz free energy, which, in a statistical
field theory, is the appropriate measure of energy to determine the
possibility of transitions between states.

The function $\xi(k) = \kT\sqrt{k^2+m^2}/\Planck$ is a nonlocal operator;
indeed, \citeSG\ prove that it is {\it anti}-local for functions in
$L^2(\RR^3)$: a function and its transform, defined by
$\tilde g(k)\mapsto\xi(k)\tilde g(k)$, can both vanish in a given region only
if the function is identically zero. Segal and Goodman relate this directly
to the Reeh-Schlieder property of a quantum field theoretic vacuum
(see, for example, \citeHaagSII53). As a simple multiplication of fourier
transforms, this operator can be understood as a convolution of $g(x)$ with
${1\over r^2}K_2(mr)$, the inverse fourier transform, up to a constant, of
$\xi(k)$. $K_2(mr)$ is a modified Bessel function which approaches zero
faster than exponentially as $r$ approaches infinity. The departure from
locality is therefore exponentially small at distances large compared to
$m^{-1}$. There is no reason to expect precisely the same behaviour as is
given by the classical heat equation, since that is a different mathematics,
but the qualitative behaviour is classically familiar. This straightforward
classical field approach characterizes the relativistic nonlocality of the
quantized Klein-Gordon field in classical terms by producing a particular
model, but it reflects Hegerfeldt's very general proof of the classical
relativistic nonlocality of quantum theory (see \citeHegerfeldt): if a
quantum system with a Hamiltonian that is bounded below is strictly
localized in a finite region, it immediately develops infinite
tails\footnote{But note that if the Hamiltonian is bounded above (i.e. the
energy is not infinite) as well as below, the quantum system cannot be
strictly localized, which is enough to preserve relativistic signal locality
(see \citeBuchholz).}. To reproduce such behaviour, a classical Hamiltonian
must certainly be relativistically nonlocal. The interesting question will
be whether this particular nonlocality is actually very unpleasant in
classical terms.

If we construct a 4-vector $k^\mu = (\omega,k)$, where
$\omega={}^{\scriptscriptstyle +}\!\sqrt{k^2+m^2}$, we can write the
Hamiltonian
$$H'[\phi] = {\kT\over\Planck}
             \int{d^3k\over(2\pi)^3} \sqrt{k^2+m^2}|\tilde\phi(k)|^2$$
as
$$H'[\phi] = {\kT\over\Planck}
             \int{d^3k\over 2\omega(2\pi)^3} 2\omega^2|\tilde\phi(k)|^2
           = {\kT\over\Planck}
             \int{d^4k\over(2\pi)^4}2\pi\delta(k^\mu k_\mu-m^2)\theta(k^0)
             2k^0 k^0|\tilde\phi(k)|^2,$$
with the additional constraint that off-shell wavenumbers have probability
zero, which enforces $\omega={}^{\scriptscriptstyle +}\!\sqrt{k^2+m^2}$ as a
constraint\footnote{It is clear that we can extend the 3-dimensional
Gaussian model into four dimensions by an arbitrary choice of $\omega$ as a
function of $k$, but the choice we have made uniquely results in a Lorentz
covariant formalism (up to a sign, of course).}. That is, $H'[\phi]$ is the
00-component of a Lorentz covariant energy-momentum tensor,
$$T^{\mu\nu}[\phi] = {\kT\over\Planck}
             \int{d^4k\over(2\pi)^4}2\pi\delta(k^\mu k_\mu-m^2)\theta(k^0)
             2k^\mu k^\nu|\tilde\phi(k)|^2,$$
which, in momentum space and for mass $m$, is the simplest possible Lorentz
covariant energy-momentum tensor. We have constructed a relativistically
covariant formalism for this classical statistical field theory, with no
preferred rest frame, despite its relativistic nonlocality. This minimalist
Lorentz covariant energy momentum tensor, as a classical equivalent of the
quantized Klein-Gordon model, deserves consideration equal to the classical
Klein-Gordon model, provided only that we are prepared to forget prejudices
against at least some forms of nonlocality.

In this covariant formulation, the probability measure over the phase at a
given wavenumber is uniform, with no correlations between the phases at
different wavenumbers, so the classical equilibrium state in one frame is
also the classical equilibrium state in another frame, and we can consider
it equivalent to the Lorentz invariant Klein-Gordon vacuum state.

When formulated in this way, the nature of the relativistic nonlocality is
not immediately clear, since there is a non-zero probability density only
for on-shell fourier modes of the classical field, none of which propagates
faster than light. In this classical statistical field theory model, the
effective nonlocality has to be taken to be due to the analytic nature of
the initial conditions.

This classical Hamiltonian can also reproduce the action of the quantized
Klein-Gordon evolution on non-vacuum states. Both for the classical model
and for the quantized Klein-Gordon model, a non-vacuum state describes a
system of probability measures different from that of the vacuum state.
A creation operator
$$a_g^\dagger=\int {d^4k\over (2\pi)^4}a^\dagger(k)\tilde g(k)$$
in quantum field theory acts on the vacuum state $\left|0\right>$, with
the Gaussian probability density
$$\rho_0(q) = {1\over\sqrt{2\pi(f,f)}}
                \exp{\left[{-q^2\over 2(f,f)}\right]},$$
where, taking $\Planck=1$, the variance $(f,f)=\sigma^2_{QFT}$ has been
expressed using the relativistically invariant inner product
$$(f,g) = \int {d^3k\over(2\pi)^3}
                {\tilde f^*(k)\tilde g(k)\over 2\sqrt{k^2+m^2}} =
          \int {d^4k\over(2\pi)^4}2\pi\delta(k^\mu k_\mu-m^2)\theta(k_0)
          {\tilde f^*(k)\tilde g(k)},$$
to give the state $a_g^\dagger\left|0\right>$, with the non-Gaussian
probability density
$$\rho_1(q) = {1\over\sqrt{2\pi(f,f)}}
                {\left[1-{|(f,g)|^2\over(f,f)(g,g)}+
                    {q^2\over(f,f)}{|(f,g)|^2\over(f,f)(g,g)}\right]}
                \exp{\left[{-q^2\over 2(f,f)}\right]}.$$
If we choose different functions for $f$ and leave $g$ fixed, then, as
the inner product $(f,g)$ changes, so the probability density varies between
$\rho_0(q)$, when $(f,g)=0$, and ${q^2\over(f,f)}\rho_0(q)$, when
$|(f,g)|^2=(f,f)(g,g)$. Note, again, that these inner products depend only
on the on-shell fourier components of $f$ and $g$. The inner product
$(f,g)$ should be carefully understood to be nonlocal: if $f(x)$ and $g(x)$
have disjoint supports, their simple inner product $\int f(x)g(x)d^3x$ will
of course be zero, but the inner product $(f,g)$ is equivalent to the simple
inner product of $f$ with $g$ in convolution with the inverse fourier
transform of $(k^2+m^2)^{-{1\over 2}}$, which will not generally be zero,
however separated the supports of $f(x)$ and $g(x)$ are, although it
{\it will} decrease faster than exponentially as the separation increases.

The probability measure $\rho_1(q)$ can be reproduced at any given time
$t_0$ within a classical statistical field theory model just as an initial
condition. $a_g^\dagger\left|0\right>$ determines a probability measure over
real functions at a time $t_0$, which we can take as a classical initial
condition. We can extend this 3-dimensional model to four dimensions by the
functional methods of \citeMorgan, which can be applied to general states
in the Fock space of the quantized Klein-Gordon model.

As further examples, the probability densities for the states
$(a_g^\dagger)^2\left|0\right>$ and $(a_g^\dagger)^3\left|0\right>$ are
$$\rho_2(q) = {1\over\sqrt{2\pi(f,f)}}
                {\left[(2-4\theta+3\theta^2)+
                  (4\theta-6\theta^2){q^2\over(f,f)}+
                  \theta^2{q^4\over(f,f)^2}\right]}
                \exp{\left[{-q^2\over 2(f,f)}\right]}$$
and
$$\rho_3(q) = {1\over\sqrt{2\pi(f,f)}}
                {\left[\matrix{(6-18\theta+27\theta^2-15\theta^3)+\cr
                  (18\theta-54\theta^2+45\theta^3){q^2\over(f,f)}+\cr
                  (9\theta^2-15\theta^3){q^4\over(f,f)^2}+
                  \theta^3{q^6\over(f,f)^3}}\right]}
                \exp{\left[{-q^2\over 2(f,f)}\right]},$$
where we have written $\theta$ for ${|(f,g)|^2\over(f,f)(g,g)}$. The above
probability densities are all even in $q$, but for the coherent
superposition $\exp(a_g^\dagger)\left|0\right>$ and for the simple linear
superposition $(ua_g^\dagger+v)\left|0\right>$, for example, terms that are
odd in $q$ appear,
$$\rho_c(q) = {1\over\sqrt{2\pi(f,f)}}
                \exp{\left[{-(q-[(f,g)+(g,f)])^2\over 2(f,f)}\right]}$$
and
$$\rho_{1,0}(q) = {1\over\sqrt{2\pi(f,f)}}
                {\left[\matrix{
        1-{|u|^2|(f,g)|^2\over(f,f)(|u|^2(g,g)+|v|^2)}+ \cr
     {q\over(f,f)}{v^*u(f,g)+u^*v(g,f)\over(|u|^2(g,g)+|v|^2)}+ \cr
     {q^2\over(f,f)}{|u|^2|(f,g)|^2\over(f,f)(|u|^2(g,g)+|v|^2)}}\right]}
                \exp{\left[{-q^2\over 2(f,f)}\right]}.$$

The closeness of this relativistically nonlocal classical statistical field
theory model to the quantized Klein-Gordon model means that we can preserve
relativistic signal locality just by restricting classical states to those
which model quantum states of bounded energy. General classical states, with
the necessary relativistically nonlocal evolution, would of course allow
relativistic signal locality to be violated, but this is no more than the
Hegerfeldt nonlocality present in quantum theory.

A quantum state, as a summary of all the correlation functions that can be
constructed for the quantized Klein-Gordon model, is not completely
reproduced by the corresponding relativistically nonlocal classical
statistical field theory model. Reflecting the modal approach to field
observables discussed in section 2, correlation functions of the form
$\left<\phi_{f_1}\phi_{f_2}...\phi_{f_n}\right>$ are equal in the two models
only when the supports of $f_i(x)$ are mutually spacelike separated. When
the supports of $f_1(x)$ and $f_2(x)$ are not spacelike separated, for
example, $\left[\phi_{f_1},\phi_{f_2}\right]\not=0$ in the quantum model,
so there is no possibility of observing a correlation function
$\left<\phi_{f_1}\phi_{f_2}\right>$ using quantum theoretically ideal
measurement devices. Classically ideal measurement devices can observe this
correlation function, but such devices would have to exhibit no quantum
fluctuations, and we have no means to achieve this. We have to explicitly
model measurement devices in a classical statistical field theory if they
are classically nonideal because they do significantly perturb the system
we wish to observe, and this ensures that measurement must be contextual in
a classical statistical field theory. We can nonetheless imagine the results
we would obtain if we did have classically ideal measurement devices, just
as we only imagine the results we would obtain if we did have quantum
theoretically ideal devices for measuring position or quantum momentum
precisely.

The quantized real Klein-Gordon model has no invariant discrete structure,
because there is no separation of the quantum state space into
superselection sectors, so the vacuum state can be continuously deformed
into the state $a^\dagger\left|0\right>$ by taking a path from $(u,v)=(0,1)$
to $(u,v)=(1,0)$. Without discrete superselection sectors, a quantum field
theory is hardly a ``quantum'' theory at all, because there is then no
invariant discrete structure in the theory. The quantized Klein-Gordon model
is better regarded as being about fields than about particles. The
superselection structures that appear in gauge theories require, however,
that some such superpositions are not allowed, so that the vacuum cannot be
continuously deformed into states that are not in the same superselection
sector. Correspondingly, a classical statistical field theory model for a
gauge theory will have to have a discrete topological structure, which will
complicate matters considerably.

\section{Is this relativistic nonlocality classically acceptable?}
A principal claim of this paper is that the classical statistical field
theory model we have constructed for the quantized Klein-Gordon model is
perfectly acceptable as classical thermal physics. Inevitably there are
differences of detail, since properties of the operators we have introduced
are not mathematically identical to properties of the heat equation, but
there is a broad similarity of faster than exponential decrease with
increasing distance.

As an approximate analogy, we can consider a detailed classical model of
sound waves in classical materials\footnote{For a close analogy, we should
consider the classical material to be in a solid phase, so that it supports
transverse waves. Quantum field theory cannot correspond {\sl exactly} to
any finite lattice (translation symmetry cannot be broken in the vacuum
state; see, for  example, \citeHaag324), but can be an effective field
theory if a lattice is sufficiently small (which would be the case for a
Planck scale lattice).} at finite temperature, taking the speed of sound as
analogous to the speed of light. In such a model, it seems that nonlocality,
relative to the `sound-cone', is largely expected. The classical wave
equation is not an adequate description of any real classical material;
there are always thermal effects, which are more-or-less described by the
heat equation. In a more adequate model, thermal effects which are outside
our control are transmitted faster than the speed of sound, reflecting the
very high thermal energy of a small number of atoms, but they allow us to
send faster-than-sound messages only under the extreme conditions of blast
waves, for example\footnote{Alternatively, consider the difficulty of
sending a message through rock at greater than the (local) speed of sound,
using only bullet-like pieces of rock.}. If electromagnetic effects were not
so accessible, the speed of sound would have been as strong a limitation on
the development of physics as the speed of light has been. Note that if we
describe the local propagation of sound using a 4-dimensional metric tensor
the conditions of a blast wave would correspond to the simultaneous
propagation of a light wave and a gravitational wave, which lies outside our
current theoretical scope in quantum physics\footnote{Taken seriously, the
analogy with sound has much to suggest to quantum gravity, because the
`sound-cone' is not microscopically defined for a lattice, and depends, for
example, on the local lattice temperature and flow. The light metric
potentially becomes an effective field rather than a description of a
fundamental geometry. Note, however, that the relativistic nonlocality of
the classical model of section 3 sits badly with the local structure of
general relativity, so that, despite the common classicality, unification
will not be helped as much as we would like by the methods of this paper.}.

The lack of localization in quantum field theory that appears very
unreasonable from the perspective of classical particle physics seems very
reasonable from the perspective of a classical thermal physics of a
continuum. We would not expect in a continuum classical thermal physics, for
example, to be able to prepare a state corresponding to a high temperature
within a finite region, while a lower temperature prevailed everywhere else,
without a thermal gradient between them, because doing so would require us
to coordinate thermal fluctuations everywhere in the universe at some
earlier time, so as to result in a thermodynamically very unlikely state
indeed. Furthermore, if we did achieve such a localized thermal state, the
heat equation would ensure that the localization would be lost instantly.

Relativistic nonlocalities of the EPR kind in the quantized Klein-Gordon
model should emerge as a consequence of the Hegerfeldt-type nonlocality we
have introduced in this classical statistical field theory model, without
any more dramatic nonlocality having to be introduced. We can perhaps expect
that the violation of Bell-type inequalities by quantized massive spin-1
models will also emerge from a similar classical model, although at a
greater theoretical distance the details of quantized massless spin-1 models
will have to be checked more carefully. Further papers will consider the
situation for quantized spin-$\Half$ models and for the standard model of
particle physics in particular. It will also be interesting to investigate
in detail whether and how the violation of Bell inequalities emerges from
just the more elementary Hegerfeldt nonlocality.

In formal terms, the classical statistical field theory model we have
constructed satisfies the relativistic principles of relativistic covariance
and relativistic signal locality, so it is acceptable as classical
relativistic physics. More as a matter of intuitive acceptability, the
(unobservable) relativistic nonlocality that is present in the theory is no
different in kind from the nonlocality that emerges in classical thermal
models that are approximately described by the heat equation, so it should
not worry us too much from a classical point of view, and it is no different
from Hegerfeldt nonlocality, so it should also not worry us too much from
a quantum point of view.

\section{Conclusion}
What is remarkable about quantum field theory from a classical perspective
is its combination of thermal and wave equations, what we can loosely call
the nonlocal and the local in classical field equations, in a single
formalism, despite the systematic, empirically motivated assertion of
relativistic signal locality. Since a relativistically covariant formalism
is possible and relativistic signal locality can be maintained despite the
relativistic nonlocality, and the relativistic nonlocality is no more than
might be expected of a classical thermal physics, it seems quite reasonable
for a classical physicist. It is perhaps enough, however, to know that such
a relativistically nonlocal classical formalism is possible, for it may well
not be mathematically more effective than the existing formalisms of quantum
field theory.

I am grateful to David Wallace and to Willem de Muynck for comments on
earlier versions of this paper.

\vfill\eject

\noindent{\Large\bf Bibliography}
\def\RefStart{\vskip 12.5pt\noindent\hangindent 30pt}

\baselineskip 13.5pt
\RefStart
   {Binney, J. J., Dowrick, N. J., Fisher, A. J., and Newman, M. E. J.
      (1993) {\sl The Theory of Critical Phenomena}
       (Oxford: Oxford University Press).}

\RefStart
   {Buchholz, D. and Yngvason, J. (1994)
      `There Are No Causality Problems for Fermi's Two-Atom System',
      {\sl Phys. Rev. Lett.} {\bf 73}, 613-616.}

\RefStart
   {Haag, R. (1996), {\sl Local Quantum Physics} (Berlin: Springer).}

\RefStart
   {Hegerfeldt, G. C. (1998) `Causality, particle localization and
         positivity of the energy',
      in: {\sl Irreversibility and Causality in Quantum Theory ---
         Semigroups and Rigged Hilbert Spaces},
      edited by A. Bohm, H.-D. Doebner and P. Kielanowski,
      Lecture Notes in Physics {\bf 504}, p. 238, (Springer: Berlin);
      quant-ph/9806036.}

\RefStart
   {Itzykson, C., and Zuber, J.-B. (1980)
       {\sl Quantum Field Theory}
       (New York: McGraw-Hill).}

\RefStart
   {Jammer, M. (1966)
      {\sl The Conceptual Development of Quantum Mechanics.}
      (New York: McGraw-Hill).}

\RefStart
   {Kaloyerou, P. N. (1996) `An ontological interpretation of boson fields',
      in: {\sl Bohmian Mechanics and Quantum Theory: An Appraisal},
      edited by J. T. Cushing {\sl et al.}, 155-167 (Kluwer: Dordrecht).}

\RefStart
   {Morgan, P. (2001) `Classical nonlocal models for states of the quantized
      real Klein-Gordon field'; quant-ph/0111027.}

\RefStart
   {Redhead, M. (1987) {\sl Incompleteness, Nonlocality, and Realism},
      (Oxford: Oxford University Press).}

\RefStart
   {Segal, I. E., and Goodman, R. W. (1965) `Anti-Locality of Certain
      Lorentz-Invariant Operators',
      {\sl Journal of Mathematics and Mechanics} {\bf 14}, 629-638.} 
\end{document}